\documentclass[aps,prd,twocolumn,groupedaddress,amssymb,eqsecnum,showpacs,epsfig]{revtex4}
\usepackage{graphicx}
\usepackage{bm}
\usepackage{dcolumn}
\usepackage{amsmath}

\def \lleq {\lower0.9ex\hbox{ $\buildrel < \over \sim$} ~}
\def \ggeq {\lower0.9ex\hbox{ $\buildrel > \over \sim$} ~}

\def \omm  {\Omega_{0 {\rm m}}}

\def \beq  {\begin{equation}}
\def \eeq  {\end{equation}}
\def \ber  {\begin{eqnarray}}
\def \eer  {\end{eqnarray}}

\def\apj{{Astroph.\@ J.\ }}

\begin{document}
\newcommand{\newc}{\newcommand}

\newc{\be}{\begin{equation}}
\newc{\ee}{\end{equation}}
\newc{\ba}{\begin{eqnarray}}
\newc{\ea}{\end{eqnarray}}
\newc{\bea}{\begin{eqnarray*}}
\newc{\eea}{\end{eqnarray*}}
\newc{\D}{\partial}
\newc{\ie}{{\it i.e.} }
\newc{\eg}{{\it e.g.} }
\newc{\etc}{{\it etc.} }
\newc{\etal}{{\it et al.}}
\newcommand{\nn}{\nonumber}
\newc{\ra}{\rightarrow}
\newc{\lra}{\leftrightarrow}
\newc{\lsim}{\buildrel{<}\over{\sim}}
\newc{\gsim}{\buildrel{>}\over{\sim}}
\title{Tension and Systematics in the Gold06 SnIa Dataset}
\author{S. Nesseris$^a$ and L. Perivolaropoulos$^b$ }
\affiliation{Department of Physics, University of Ioannina, Greece
\\ $^a$ e-mail: me01629@cc.uoi.gr, $^b$ e-mail:
leandros@uoi.gr}
\date {\today}

\begin{abstract}
The Gold06 SnIa dataset recently released in astro-ph/0611572
consists of five distinct subsets defined by the group or
instrument that discovered and analyzed the corresponding data.
These subsets are: the SNLS subset (47 SnIa), the HST subset (30
SnIa), the HZSST subset (41 SnIa), the SCP subset (26 SnIa) and
the Low Redshift (LR) subset (38 SnIa). These subsets sum up to
the 182 SnIa of the Gold06 dataset. We use Monte-Carlo simulations
to study the statistical consistency of each one of the above
subsets with the full Gold06 dataset. In particular, we compare
the best fit $w(z)$ parameters $(w_0,w_1)$ obtained by subtracting
each one of the above subsets from the Gold06 dataset (subset
truncation), with the corresponding best fit parameters
$(w^r_0,w^r_1)$ obtained by subtracting the same number of {\it
randomly selected} SnIa from the same redshift range of the Gold06
dataset (random truncation). We find that the probability for
$(w^r_0,w^r_1)=(w_0,w_1)$ is large for the  Gold06 minus SCP
(Gold06-SCP) truncation but is less than $5\%$ for the
Gold06-SNLS, Gold06-HZSST and Gold06-HST truncations. This result
implies that the Gold06 dataset is not statistically homogeneous.
By comparing the values of the best fit $(w_0,w_1)$ for each
subset truncation we find that the tension among subsets is such
that the SNLS and HST subsets are statistically consistent with
each other and `pull' towards $\Lambda$CDM $(w_0=-1,w_1=0)$ while
the HZSST subset is statistically distinct and strongly `pulls'
towards a varying $w(z)$ crossing the line $w=-1$ from below
$(w_0<-1,w_1>0)$. We also isolate six SnIa that are mostly
responsible for this behavior of the HZSST subset.
\end{abstract}
\pacs{98.80.Es,98.65.Dx,98.62.Sb}
\maketitle

\section{Introduction}
Current cosmological observations show strong evidence that we
live in a spatially flat universe \cite{Spergel:2003cb} with low
matter density \cite{Tegmark:2003ud}  that is currently undergoing
accelerated cosmic expansion. The most direct indication for the
current accelerating expansion comes from the accumulating type Ia
supernovae (SnIa) data
\cite{hzsst,scp,lr,snobs,Astier:2005qq,Riess:2004nr,Riess:2006fw}
which provide a detailed form of the recent expansion history of
the universe.

This accelerating expansion has been attributed to a dark energy
component with negative pressure which can induce repulsive
gravity and thus cause accelerated expansion (for recent reviews
see
\cite{Padmanabhan:2006cj,Copeland:2006wr,Perivolaropoulos:2006ce,Alcaniz:2006ay,Sahni:2006pa,Uzan:2006mf,Polarski:2006ut})
The simplest and most obvious candidate for this dark energy is
the cosmological constant $\Lambda$ \cite{Sahni:1999gb} with
equation of state $w={p / \rho}=-1$. This model however raises
theoretical problems related to the fine tuned value required for
the cosmological constant. These difficulties have lead to a large
variety of proposed models where the dark energy component evolves
with time usually due to an evolving scalar field (quintessence)
which may be minimally \cite{quintess} or non-minimally
\cite{modgrav} coupled to gravity. Alternatively, more general
modified gravity theories\cite{Nojiri:2006ri} have also been
proposed based on $f(R)$
theories\cite{Chiba:2003ir,Nojiri:2003ft,Nojiri:2006be} (for a
debate on the issue see \cite{Amendola:2006kh}),
braneworlds\cite{Aref'eva:2005fu,Lazkoz:2006gp,Bogdanos:2006dt,Cai:2005qm},
Gauss-Bonnet dark energy\cite{Nojiri:2005vv}, holographic dark
energy\cite{Zhang:2005yz} etc. The main prediction of the
dynamical models is the evolution of the dark energy density
parameter $\Omega_X(z)$. Combining this prediction with the prior
assumption for the matter density parameter $\Omega_{0m}$, the
predicted expansion history $H(z)$ is obtained as \be H(z)^2 =
H_0^2 [\Omega_{0m} (1+z)^3 + \Omega_X (z)] \ee The dark energy
density parameter is usually expressed as \be \Omega_X
(z)=\Omega_{0X} e^{3\int_0^z \frac{dz'}{1+z'}(1+w(z'))} \ee where
$w(z)$ is related to $H(z)$ by
\cite{Saini:1999ba,Nesseris:2004wj,Huterer:2000mj}  \be
\label{wz3} w(z)={{{2\over 3} (1+z) {{d \ln H}\over {dz}}-1} \over
{1-({{H_0}\over H})^2 \Omega_{0m} (1+z)^3}} \ee If the dark energy
can be described as an ideal fluid with conserved energy momentum
tensor $T^{\mu \nu}=diag(\rho, p, p, p) $ then the above parameter
$w(z)$ is identical with the equation of state parameter of dark
energy \be w(z)=\frac{p(z)}{\rho(z)} \ee Independently of its
physical origin, the parameter $w(z)$ is an observable derived
from $H(z)$ (with prior knowledge of $\Omega_{0m}$) and is usually
used to compare theoretical model predictions with observations.

The two most reliable and robust SnIa datasets existing at present
are the Gold dataset \cite{Riess:2006fw} (hereafter Gold06) and
the Supernova Legacy Survey (SNLS) \cite{Astier:2005qq} dataset.
The Gold dataset compiled by Riess et. al. is a set of 182
supernova data from various sources analyzed in a consistent and
robust manner with reduced calibration errors arising from
systematics. It contains 119 points from previously published data
\cite{Riess:2004nr} (hereafter Gold04) plus 16 points with
$0.46<z<1.39$ discovered recently by the Hubble Space Telescope
(HST). It also incorporates 47 points ($0.25<z<1$) from the first
year release of the SNLS dataset \cite{Astier:2005qq} out of a
total of 73 distant SnIa. Some supernovae were
excluded\cite{Riess:2006fw} due to highly uncertain color
measurements, high extinction $A_V>0.5$ and a redshift cut $c z<
7000km/s$ or $z<0.0233$, to avoid the influence of a possible
local ``Hubble Bubble", so as to define a high-confidence
subsample. In addition, a single algorithm (MLCS2k2) was applied
to estimate all the SnIa distances (including those originating
from SNLS) thus attempting to minimize the non-uniformities of the
dataset.

The total of 182 SnIa included in the Gold06 dataset can be
grouped into five subsets according to the search
teams/instruments that discovered them. These subsets are shown in
Table I. A detailed table of all the data used in our analysis and
their subset origin is shown in the Appendix. Notice that the
early data of the Gold06 dataset were obtained mainly in the 90's
and consist of the High z Supernova Search Team (HZSST) subset,
the Supernova Cosmology Project (SCP) subset and the Low Redshift
(LR) subset.

The above observations provide the apparent magnitude $m(z)$ of
the supernovae at peak brightness after implementing correction
for galactic extinction, K-correction and light curve
width-luminosity correction. The resulting apparent magnitude
$m(z)$ is related to the luminosity distance $d_L(z)$ through \be
m_{th}(z)={\bar M} (M,H_0) + 5 log_{10} (D_L (z)) \label{mdl} \ee
where in a flat cosmological model \be D_L (z)= (1+z) \int_0^z
dz'\frac{H_0}{H(z';a_1,...,a_n)} \label{dlth1} \ee is the Hubble
free luminosity distance ($H_0 d_L/c$), $a_1,...,a_n$ are
theoretical model parameters and ${\bar M}$ is the magnitude zero
point offset and depends on the absolute magnitude $M$ and on the
present Hubble parameter $H_0$ as \ba
{\bar M} &=& M + 5 log_{10}(\frac{c\; H_0^{-1}}{Mpc}) + 25= \nn \\
&=& M-5log_{10}h+42.38 \label{barm} \ea The parameter $M$ is the
absolute magnitude which is assumed to be constant after the above
mentioned corrections have been implemented in $m(z)$.

\vspace{0pt}
\begin{table}[t!]
\begin{center}
\caption{The subsets of the Gold06 dataset (see also
\cite{snweb}). \label{table1}}
\begin{tabular}{ccccc}
\hline
\hline\\
\vspace{6pt} \textbf{Subsets} \hspace{7pt}& \textbf{Total} \hspace{7pt}& \textbf{Redshift Range} \hspace{7pt}& \textbf{Years of discovery} \hspace{7pt}& \textbf{Ref}. \\
\vspace{6pt}  SNLS           & 47       \hspace{7pt}& $0.25\leq z \leq 0.96$ \hspace{7pt}& 2003-2004  \hspace{7pt}& \cite{Astier:2005qq} \\
\vspace{6pt}  HST            & 30       \hspace{7pt}& $0.46\leq z\leq 1.76$ \hspace{7pt}& 1997-2005  \hspace{7pt}& \cite{Riess:2006fw} \\
\vspace{6pt}  HZSST          & 41       \hspace{7pt}& $0.28\leq z\leq 1.20$ \hspace{7pt}& 1995-2001  \hspace{7pt}& \cite{hzsst} \\
\vspace{6pt}  SCP            & 26       \hspace{7pt}& $0.17\leq z\leq 0.86$ \hspace{7pt}& 1995-2000  \hspace{7pt}& \cite{scp} \\
\vspace{6pt}  LR             & 38       \hspace{7pt}& $0.024\leq z\leq 0.12$ \hspace{7pt}& 1990-2000  \hspace{7pt}& \cite{lr} \\
  \hline \hline
\end{tabular}
\end{center}
\end{table}

The data points of the Gold06 dataset are given after the
corrections have been implemented, in terms of the distance
modulus \be \mu_{obs}(z_i)\equiv m_{obs}(z_i) - M \label{mug}\ee
The theoretical model parameters are determined by minimizing the
quantity \be \chi^2 (a_1,...,a_n)= \sum_{i=1}^N
\frac{(\mu_{obs}(z_i) - \mu_{th}(z_i))^2}{\sigma_{\mu \; i}^2 +
\sigma_{v\; i}^2 } \label{chi2} \ee where $\sigma_{\mu \; i}^2$
and $\sigma_{v\; i}^2$ are the errors due to flux uncertainties
and peculiar velocity dispersion respectively. These errors are
assumed to be gaussian and uncorrelated. The theoretical distance
modulus is defined as \be \mu_{th}(z_i)\equiv m_{th}(z_i) - M =5
log_{10} (D_L (z)) +\mu_0 \label{mth} \ee where \be \mu_0= 42.38 -
5 log_{10}h \label{mu0}\ee and $\mu_{th}(z_i)$ also depends on the
parameters $a_1,...,a_n$ used in the parametrization of $H(z)$ in
equation (\ref{dlth1}).

The parametrization used in our analysis is the CPL
parametrization \cite{Chevallier:2000qy,Linder:2002et} \ba
w(z)&=&w_0+w_1 \frac{z}{1+z} \label{lda}
\\ H^2 (z)&=&H_0^2 [ \Omega_{0m} (1+z)^3
+ \nn \\ +(1-\Omega_{0m})
(1&+&z)^{3(1+w_0+w_1)}e^{3w_1[1/(1+z)-1]}]\ea with a prior of the
matter density parameter $\omm= 0.28$ (as in Ref.
\cite{Riess:2006fw}), assuming flatness, according to the methods
described in detail in Ref. \cite{Lazkoz:2005sp,Nesseris:2005ur}.

\begin{figure*}[t!]
\rotatebox{0}{\resizebox{1\textwidth}{!}{\includegraphics{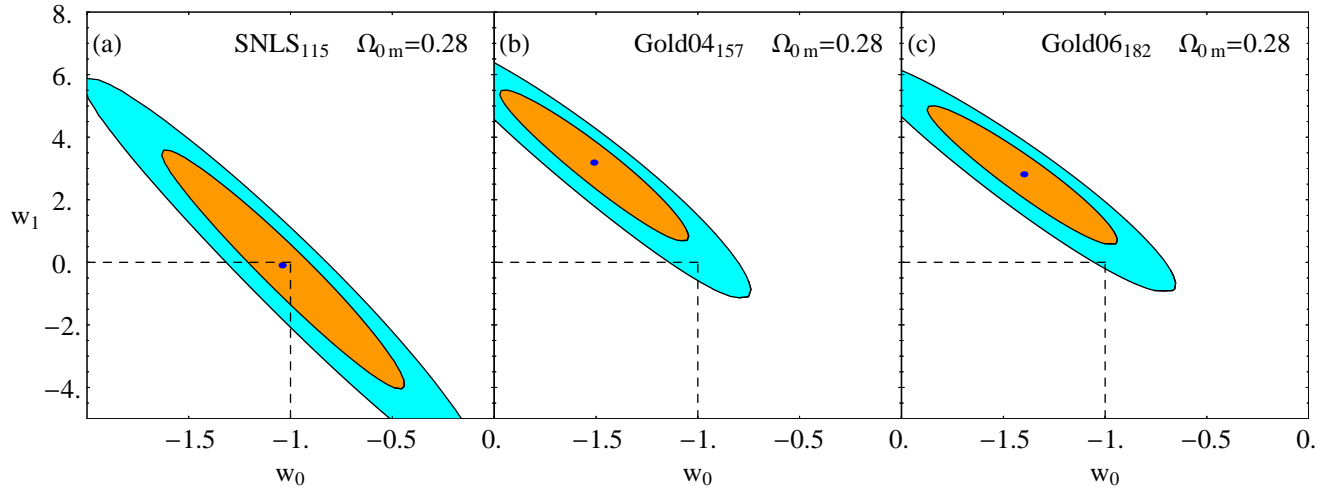}}}
\vspace{0pt}{ \caption{Maximum likelihood fits of the CPL
parametrization (\ref{lda}) to the SNLS (a) Gold04 (b) and Gold06
(c) datasets. The $2\sigma$ tension between the Gold and SNLS
remains with the new Gold06 dataset despite the improved
filtering, calibration and data extension.}} \label{fig1}
\end{figure*}

The previous version of the Gold sample \cite{Riess:2004nr}
(Gold04) had been shown to be in mild ($2\sigma$) tension with the
SNLS dataset \cite{Nesseris:2005ur,Jassal:2005qc}. While the
Gold04 mildly favored an evolving dark energy equation of state
parameter $w(z)$ (crossing the phantom divide line w=-1) over the
cosmological constant ($\Lambda$CDM) at almost $2\sigma$ level
\cite{Alam:2004jy,Wang:2006ts,Daly:2006ax,Huterer:2004ch,Shafieloo:2005nd,Lazkoz:2005sp},
the SNLS data had shown no such trend and provided
\cite{Nesseris:2005ur} a best fit $w(z)$ very close to $w=-1$
($\Lambda$CDM). The trend towards phantom divide crossing can not
be explained in the context of minimally coupled quintessence and
could be viewed as an indication for more exotic
models\cite{Gannouji:2006jm,Guo:2006pc,Kujat:2006vj,Zhang:2006at,Nesseris:2006hp,Briscese:2006xu,Kahya:2006hc,McInnes:2005vp}.
This mild tension could have been attributed to systematic errors
due eg to the different algorithm used in the analysis of the two
datasets. The new version of the Gold sample however, (Gold06)
involves an improved uniform analysis and incorporates a large
part of the SNLS sample. Thus there could have been an
anticipation that the mild tension with SNLS would be ameliorated
or even disappear. As shown in Fig. 1 however, this anticipation
has not been fulfilled (see also \cite{Alam:2006kj,Gong:2006gs}).

The mild (almost $2\sigma$) tension between the Gold04 and the
SNLS samples (Figs. 1a and 1b) has not decreased by using the
Gold06 sample (Fig. 1c)! The investigation of the origin of this
tension  and the statistical uniformity of the Gold06 dataset
consist the main focus of the present paper.

\section{Tension in the Gold06 Dataset}
The 182 SnIa included in the Gold06 dataset originate mainly from
the search teams/instruments shown in Table I. The low redshift
subset (LR) is a mixture of various early SnIa by different groups
and instruments but we consider it as a single subset because
otherwise we would have to increase the number of subsets beyond a
reasonable number.

In order to investigate the statistical uniformity of the Gold06
dataset and also the origin of the tension with the SNLS, we have
decomposed the Gold06 dataset into the subsamples of Table I and
constructed new datasets by subtracting each one (or two) of the
subsets from the full Gold06 dataset. We thus obtained the
following six subset truncations:
\begin{enumerate} \item $182_{G06}-47_{SNLS}-30_{HST}$ \item
$182_{G06}-47_{SNLS}$ \item $182_{G06}-30_{HST}$ \item
$182_{G06}-26_{SCP}$ \item $182_{G06}-41_{HZSST}$ \item
$182_{G06}-41_{HZSST}-26_{SCP}$ \end{enumerate} We did not
consider the subset $182_{G06}-38_{LR}$ with low redshift
truncation because the $LR$ subset is not uniform and also because
subtracting it can not be associated with a corresponding random
truncation in the same low redshift range (the range $z<0.124$ is
spanned completely by the LR subset). We then addressed the
following two questions:
\begin{itemize} \item How do the best fit $(w_0,w_1)$ values for
each of the six truncations compare with the corresponding best
fit value of the full Gold06 dataset? \item How do the best fit
$(w_0,w_1)$ values for each of the six truncations compare with
the corresponding best fit value of a random truncation of the
full Gold06 dataset made in the same redshift range as that of the
subtracted subset? \end{itemize} The answer to the first question
is provided in Fig. 2 where we show the best fit values
$(w_0,w_1)$ for each one of the above six truncations. Notice that
the two multiple truncations: $182_{G06}-41_{HZSST}-26_{SCP}$
(point 1) and $182_{G06}-47_{SNLS}-30_{HST}$ (point 6) correspond
to more extreme best fit values of $(w_0,w_1)$. The best fit
$(w_0,w_1)$ of the Gold06 dataset along with its $1\sigma$ and
$2\sigma$ contours is also shown in Fig. 2 (point 0).

The following comments can be made on the basis of Fig. 2:
\begin{itemize}\item The truncation $182_{G06}-26_{SCP}$
leaves the best fit $(w_0,w_1)$ of the Gold06 dataset practically
unchanged \item No single subset truncation is able to shift the
best fit $(w_0,w_1)$ values beyond the $1\sigma$ contours of the
Gold06 dataset. \clearpage

\begin{figure}
\hspace{0pt}\rotatebox{0}{\resizebox{.5\textwidth}{!}{\includegraphics{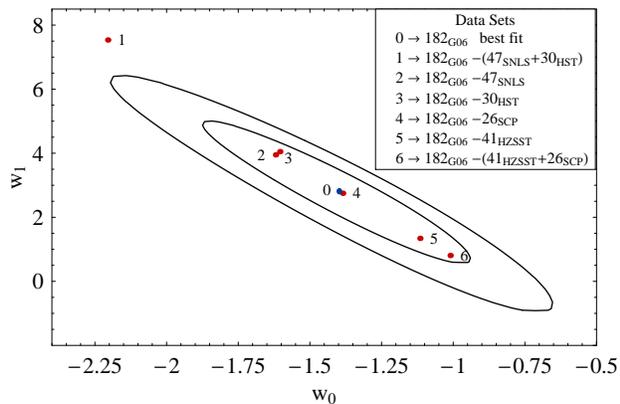}}}
\vspace{-20pt}{\caption{The $1 \sigma-2 \sigma$ $\chi^2$
confidence region ellipses in the $w_{0}-w_{1}$ plane based on
parametrization (\ref{lda}) for the Gold06 dataset and $\omm=
0.28$. Superposed are the best fit parameter values for each one
of the truncations 1-6 of the Gold06 dataset.}} \label{fig2}
\end{figure}

\item All the subset truncations (except $182_{G06}-26_{SCP}$)
systematically shift the best fit $(w_0,w_1)$ along the major axis
of the $\chi^2$ ellipse. In particular for $182_{G06}-30_{HST}$
and $182_{G06}-47_{SNLS}$ the best fit is left mainly under the
influence of $HZSST$ and is shifted along the major axis, away
from $\Lambda$CDM towards an evolving $w(z)$ crossing the line
$w=-1$ ($w_0<-1$, $w_1>0$). On the other hand for
$182_{G06}-41_{HZSST}$ the best fit $(w_0,w_1)$ is left under the
influence of $HST$ and $SNLS$ and is shifted towards $\Lambda$CDM.
This implies that the subsets $HST$ and $SNLS$ favor $\Lambda$CDM
while the subset $HZSST$ favors an evolving $w(z)$ crossing the
phantom divide $w=-1$. This result is further amplified by the
behavior of the multiple truncations
$182_{G06}-41_{HZSST}-26_{SCP}$ (further shifted towards
$\Lambda$CDM) and $182_{G06}-47_{SNLS}-30_{HST}$ (strongly shifted
towards a varying $w(z)$ crossing the phantom divide $w=-1$ at a
level more than $2\sigma$ (see Fig. 2)).\end{itemize}

Based on the above comments we conclude that the answer to the
first question stated above can be summarized as follows: The best
fit $(w_0,w_1)$ values for each of the four single set truncations
2-5 do not differ more than $1\sigma$ from the best fit
corresponding values of the Gold06 dataset but they show distinct
trends which are characteristic for each one of the truncations.

A separate question (related to the second question stated above)
is the question of statistical consistency between each subset
truncation and the full Gold06 dataset. To address this question
we compare the best fit value of $(w_0,w_1)$ for each subset
truncation with a large number (500) of corresponding random
truncations of the Gold06 dataset. The random truncations involve
random subtractions of the same number of SnIa and in the same
redshift range as the subset truncation. These random truncations
can be used to obtain the $1\sigma$ range for the

\vspace{0pt}
\begin{table}[h!]
\begin{center}
\caption{The six subset truncations of Fig 3. \label{table2}}
\begin{tabular}{cccc}
\hline
\hline\\
\vspace{6pt}  Dataset                   & $^{w_0}_{w_1}$ & $^{{w}^r_0}_{{w}^r_1}$ (MC) & $\frac{w-{\bar w}^r}{\sigma_{w^r}}$  \\
\vspace{6pt}  $182 -47_{SNLS}-30_{HST}$ & $^{w_0=-2.21}_{w_1=\;\;\; 7.53}$ \hspace{7pt}& $^{{w}^r_0=-1.40 \pm 0.22}_{{w}^r_1= \;\;\; 2.83 \pm 1.30}$ \hspace{7pt}& $~-3.7\sigma$ \\
\vspace{6pt}  $182 -47_{SNLS}$          & $^{w_0=-1.62}_{w_1=\;\;\; 3.95}$ \hspace{7pt}& $^{{w}^r_0=-1.38 \pm 0.12}_{{w}^r_1=\;\;\; 2.67 \pm 0.57}$ \hspace{7pt}& $~-2.2\sigma$  \hspace{7pt}\\
\vspace{6pt}  $182 -30_{HST}$           & $^{w_0=-1.60}_{w_1=\;\;\; 4.05}$ \hspace{7pt}& $^{{w}^r_0=-1.36 \pm 0.10}_{{w}^r_1=\;\;\; 2.60 \pm 0.64}$ \hspace{7pt}& $~-2.4\sigma$  \hspace{7pt}\\
\vspace{6pt}  $182 -26_{SCP}$           & $^{w_0=-1.39}_{w_1=\;\;\; 2.75}$ \hspace{7pt}& $^{{w}^r_0=-1.40 \pm 0.08}_{{w}^r_1=\;\;\; 2.79 \pm 0.38}$ \hspace{7pt}& $~+0.2\sigma$  \hspace{7pt}\\
\vspace{6pt}  $182 -41_{HZSST}$         & $^{w_0=-1.12}_{w_1=\;\;\; 1.34}$ \hspace{7pt}& $^{{w}^r_0=-1.40 \pm 0.11}_{{w}^r_1=\;\;\; 2.80 \pm 0.55}$ \hspace{7pt}& $~+2.7\sigma$  \hspace{7pt}\\
\vspace{6pt}  $182 -41_{HZSST}-26_{SCP}$  & $^{w_0=-1.01}_{w_1=\;\;\; 0.81}$ \hspace{7pt}& $^{{w}^r_0=-1.39 \pm 0.15}_{{w}^r_1=\;\;\; 2.75 \pm 0.73}$ \hspace{7pt}& $~+2.6\sigma$  \hspace{7pt}\\
  \hline \hline
\end{tabular}
\end{center}
\end{table}

\noindent expected values of the best fit $(w^r_0,w^r_1)$ of the
randomly truncated Gold06 dataset.

If the best fit values $(w_0,w_1)$ of the subset truncation is
within the $1\sigma$ range of the best fit values $(w^r_0,w^r_1)$
of the random truncation then the considered subset truncation is
a typical truncation representative of the Gold06 dataset and
statistically consistent with it. If on the other hand $(w_0,w_1)$
differs by $2\sigma$ or more from the mean best fit values $({\bar
w}^r_0,{\bar w}^r_1)$ of the random truncation then the considered
subset truncation is not a typical truncation and is
systematically different from the full dataset. We have
implemented the above comparison for the six subset truncations
referred above and the results are shown in Table II and in Fig.
3.

The following comments can be made on the basis of Table II and
Fig. 3: \begin{itemize} \item The $SCP$ is a typical,
statistically consistent subset of the Gold06 dataset because its
truncation does not significantly alter the statistical properties
of the Gold06 dataset. In particular the best fit $(w_0,w_1)$
value of the $182_{G06}-26_{SCP}$ truncation differs only by
$0.2\sigma$ from the corresponding mean random truncation best fit
$({\bar w}^r_0,{\bar w}^r_1)$ which involves random subtraction of
the same number of SnIa from the same redshift range as the SCP
subset. \item The other five subsets considered in Fig. 3 are not
typical subsets of the Gold06 dataset. The best fit $(w_0,w_1)$
values of the truncations considered in Fig. 3 differ by more than
$2\sigma$ from the mean best fit values $({\bar w}^r_0,{\bar
w}^r_1)$ of the corresponding random truncations.

\begin{figure*}[t!]
\rotatebox{0}{\resizebox{1\textwidth}{!}{\includegraphics{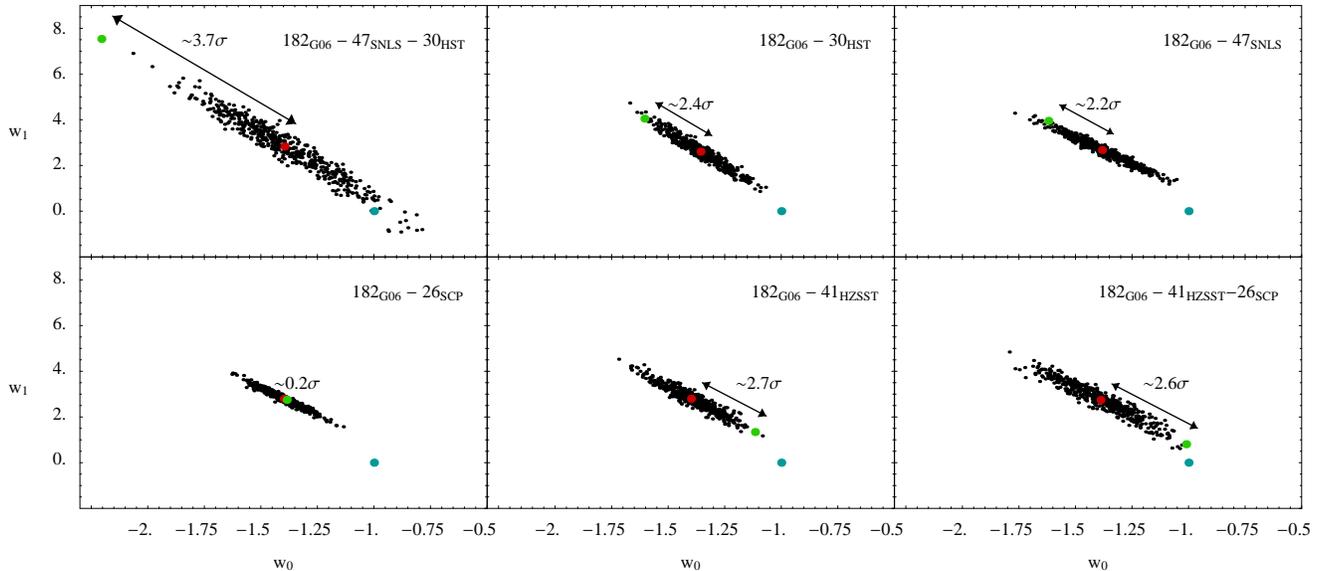}}}
\vspace{0pt}{ \caption{Comparison of the best fit parameters to
the subsample truncations 1-6 with corresponding random
truncations of the Gold06 dataset. In all truncation cases (except
of the SCP truncation) the best fit parameter values are shifted
(in different directions) by more than $2\sigma$ from the mean
random truncation values. The point corresponding to $\Lambda$CDM
($w_0=-1,w_1=0$) is also shown.}} \label{fig3}
\end{figure*}

\item An extreme case is the truncation
$182_{G06}-47_{SNLS}-30_{HST}$ whose best fit values are $3.7
\sigma$ away from the corresponding mean best fit values of a
random truncation! This implies that the combination of the
$38_{LR}+41_{HZSST}+26_{SCP}$ which is left over from the
truncation $182_{G06}-47_{SNLS}-30_{HST}$ strongly favors an
evolving $w(z)$ and is statistically inconsistent with the Gold06
dataset. This result is consistent with Fig. 2 which also shows
that best fit $(w_0,w_1)$ of the truncation
$182_{G06}-47_{SNLS}-30_{HST}$ is about $3\sigma$ away from the
Gold06 best fit! \item The $SNLS$ and $HST$ subsets are
statistically very similar to each other (with a trend towards
$\Lambda$CDM) even though they are both significantly different
(more that $2\sigma$) from the corresponding random truncations of
Gold06 (see also Fig. 2).\item Both Figs 2 and 3 indicate that the
trend towards $\Lambda$CDM increases for more recent ($HST$ and
$SNLS$) data while earlier data ($HZSST$ and $SCP$) seem to favor
and evolving $w(z)$.
\end{itemize}

The above results can also be verified by considering the `pure'
Gold06 dataset which does not include the 47 SnIa of SNLS. This
dataset (Gold06p) consists of 135 SnIa and is essentially a
filtered version of the Gold04 dataset with the addition of the 16
SnIa with $0.46<z<1.39$ discovered recently by the HST. The best
fit parameter values for the Gold06p dataset are somewhat shifted
in the direction of varying $w(z)$ compared to the full Gold06
(compare Figs. 2 and 4) as expected since SNLS favors
$\Lambda$CDM. As shown in Fig. 4 and Table III, the effect of each
subset truncation in this case is more prominent due to the
smaller number of points in the Gold06p dataset.

\begin{figure}[h!]
\hspace{0pt}\rotatebox{0}{\resizebox{.5\textwidth}{!}{\includegraphics{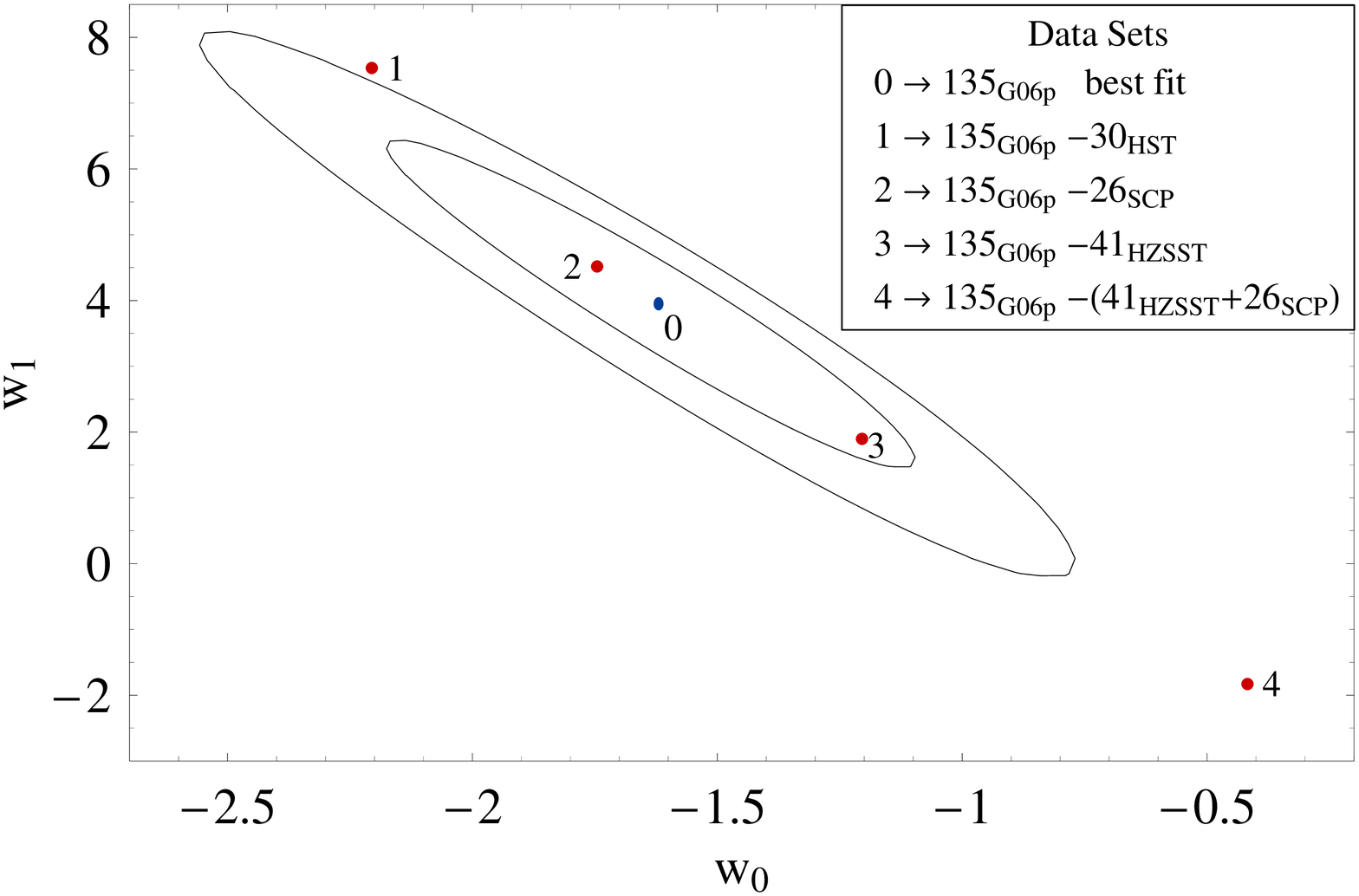}}}
\vspace{-20pt}{\caption{The $1 \sigma-2 \sigma$ $\chi^2$
confidence region ellipses in the $w_{0}-w_{1}$ plane based on
parametrization (\ref{lda}) for the Gold06p dataset and $\omm=
0.28$. Superposed are the best fit parameter values for each one
of four truncations of the Gold06p dataset. The best fit
parameters for the truncation $135_{G06p}-41_{HZSST}-26_{SCP}$ are
shifted by about $3\sigma$ from corresponding Gold06p best fit
values in the direction of $\Lambda$CDM.}} \label{fig4}
\end{figure}

\vspace{0pt}
\begin{table}[b!]
\begin{center}
\caption{The four subset truncations of Fig 5. \label{table3}}
\begin{tabular}{cccc}
\hline
\hline\\
\vspace{6pt}  Dataset                   & $^{w_0}_{w_1}$ & $^{{w}^r_0}_{{w}^r_1}$ (MC) & $\frac{w-{\bar w}^r}{\sigma_{w^r}}$  \\
\vspace{6pt}  $135 -30_{HST}$           & $^{w_0=-2.21}_{w_1=\;\;\; 7.53}$ \hspace{7pt}& $^{{w}^r_0=-1.63 \pm 0.17}_{{w}^r_1=\;\;\; 3.98 \pm 0.97}$ \hspace{7pt}& $~-3.6\sigma$  \hspace{7pt}\\
\vspace{6pt}  $135 -26_{SCP}$           & $^{w_0=-1.75}_{w_1=\;\;\; 4.52}$ \hspace{7pt}& $^{{w}^r_0=-1.63 \pm 0.17}_{{w}^r_1=\;\;\; 3.91 \pm 0.75}$ \hspace{7pt}& $~-0.8\sigma$  \hspace{7pt}\\
\vspace{6pt}  $135 -41_{HZSST}$         & $^{w_0=-1.20}_{w_1=\;\;\; 1.90}$ \hspace{7pt}& $^{{w}^r_0=-1.60 \pm 0.21}_{{w}^r_1=\;\;\; 3.76 \pm 0.95}$ \hspace{7pt}& $~+1.9\sigma$  \hspace{7pt}\\
\vspace{6pt}  $135 -41_{HZSST}-26_{SCP}$  & $^{w_0=-0.42}_{w_1=- 1.83}$ \hspace{7pt}& $^{{w}^r_0=-1.67 \pm 0.37}_{{w}^r_1=\;\;\; 3.95 \pm 1.55}$ \hspace{7pt}& $~+3.7\sigma$  \hspace{7pt}\\
  \hline \hline
\end{tabular}
\end{center}
\end{table}

For example, the $135_{G06p}-41_{HZSST}-26_{SCP}$ truncation
shifts the best fit parameter values of the Gold06p by about
$3\sigma$ in the direction of $\Lambda$CDM (and beyond it) while
the shift with respect to the random truncations of Gold06p is
$3.7\sigma$ (Fig. 5). The corresponding shifts with respect to the
Gold06 dataset were about $1\sigma$ and $2.6\sigma$ respectively
(Figs. 2 and 3).

\begin{figure*}[t!]
\rotatebox{0}{\resizebox{1\textwidth}{!}{\includegraphics{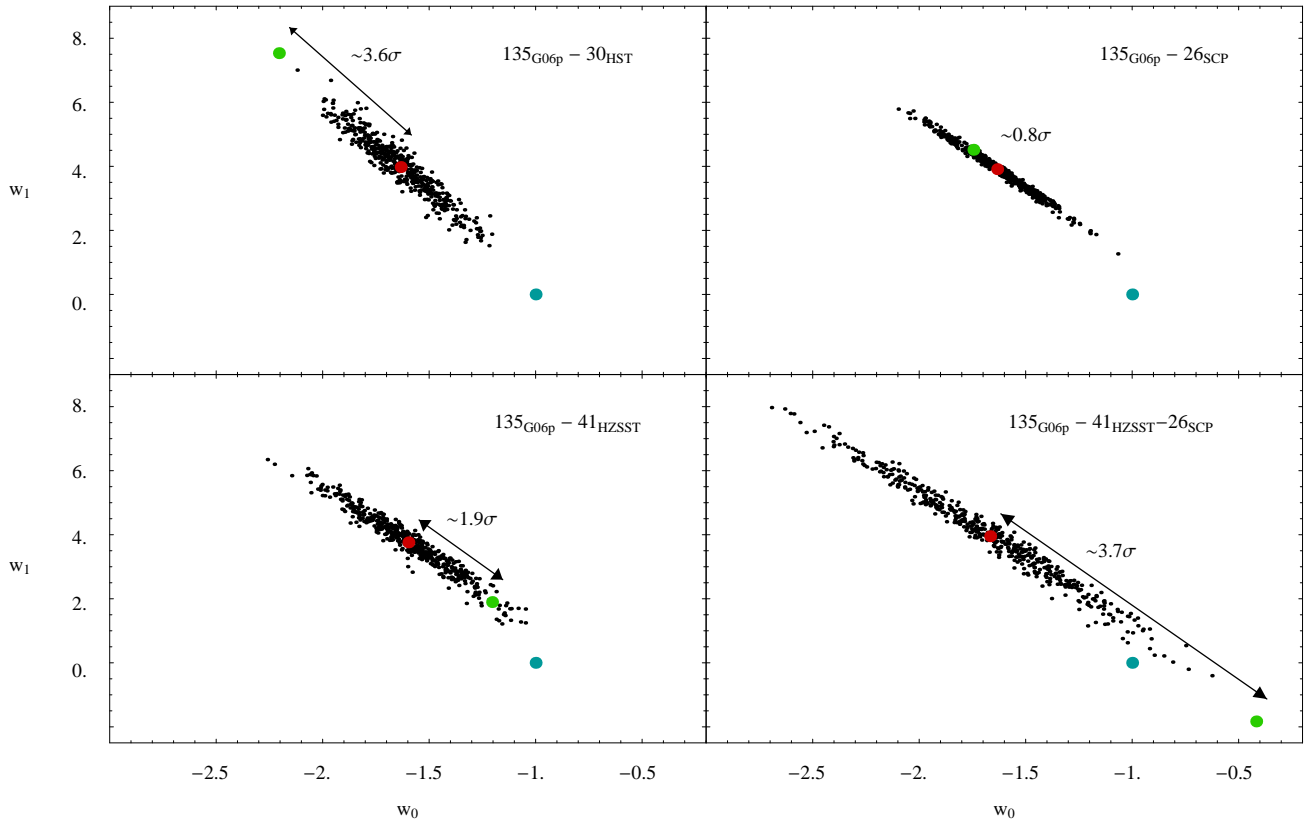}}}
\vspace{0pt}{ \caption{Comparison of the best fit parameters to
the subsample truncations shown in Table III with corresponding
random truncations of the Gold06p dataset. In all truncation cases
(except of the SCP truncation) the best fit parameter values are
shifted (in different directions) by more than $2\sigma$ from the
mean random truncation values. The best fit parameter shift of the
$135_{G06p}-41_{HZSST}-26_{SCP}$ is $3.7\sigma$ compared to the
corresponding random truncation. The point corresponding to
$\Lambda$CDM ($w_0=-1,w_1=0$) is also shown.}} \label{fig5}
\end{figure*}

\section{Discussion-Conclusion}
The fact that more recent SnIa data (HST and SNLS) seem to favor
$\Lambda$CDM significantly more than earlier data (HZSST) makes it
possible that earlier data may be more prone to systematic errors.
It is therefore interesting to identify a small subset of SnIa
from the HZSST data that is mostly responsible for the trend of
HZSST towards an evolving $w(z)$. We have isolated the group of
SnIa in the HZSST subset whose distance modulus differs by more
than $1.8\sigma$ from the $\Lambda$CDM predictions ($\omm=0.28$).
The group which consists of just six SnIa is also significantly
responsible for the trend of the HZSST subset towards an evolving
$w(z)$. These SnIa are: (SN99Q2, SN00ee, SN00ec, SN99S, SN01fo,
SN99fv). The shifted best fit parameter values $(w_0,w_1)$ due to
these six SnIa data truncation are shown in Fig. 6a superposed on
a Monte-Carlo simulation of corresponding random 6 point
truncations to the HZSST subset. We anticipate that the possible
systematic errors that lead to the distinct behavior of the HZSST
subset are maximal for these six SnIa and it may be easier to
identify them and correct them in this set of six SnIa.
Alternatively these 6 SnIa could be discarded from the Gold06
dataset as outliers in an effort to improve its statistical
uniformity and bring it to line with the more recent data.

\begin{figure*}[t!]
\rotatebox{0}{\resizebox{1\textwidth}{!}{\includegraphics{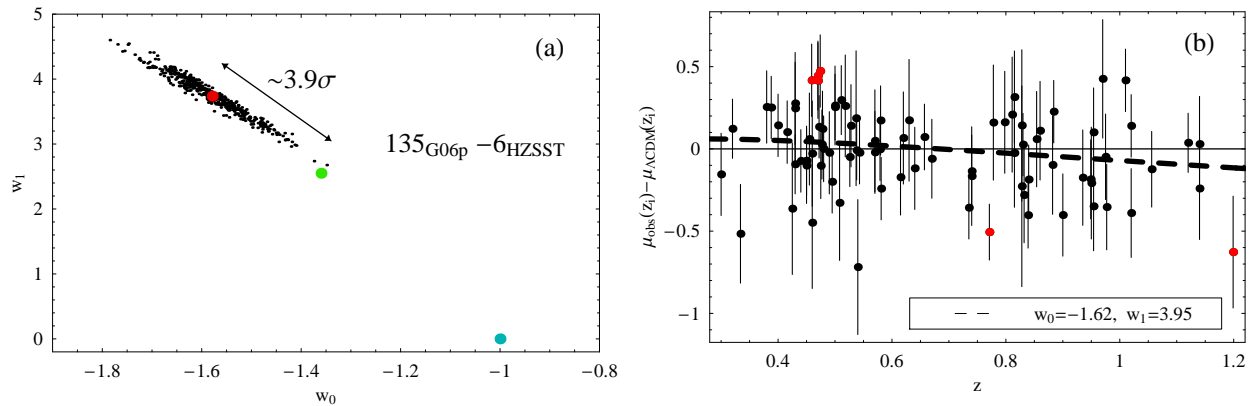}}}
\vspace{0pt}{ \caption{a. The best fit parameter values for the
Gold06p dataset for random 6 point truncations from the HZSST
subset. The parameter shift is maximized at $~3.9\sigma$ when the
following six points are truncated: (SN99Q2, SN00ee, SN00ec,
SN99S, SN01fo, SN99fv) (red dot). These are also the points whose
distance modulus differs by more than $1.8\sigma$ from the
$\Lambda CDM$ predictions. b. The best fit distance modulus
(dashed line) relative to $\Lambda$CDM and the data of the Gold06
dataset in the redshift range of the HZSST subset. The six points
of the HZSST subset which differ from $\Lambda CDM$ by more than
$1.8\sigma$ are colored in red. They are also the most favorable
points for an evolving $w(z)$.}} \label{fig6}
\end{figure*}

A visual display of the six datapoints (points in red) compared to
other datapoints is shown in Fig. 6b where we show the distance
modulus relative to $\Lambda$CDM ($\omm=0.28$) of the Gold06 data
in the redshift range of the HZSST subset. In the same plot we
show (thick dashed line) the distance modulus corresponding to the
best fit values $(w_0=-1.62,w_1=3.95)$ obtained from the Gold06p
data (dashed line) indicating that all of the six red datapoints
strongly favor the best fit $w(z)$ over $\Lambda$CDM.

In conclusion we have demonstrated that despite the careful
filtering and the improved calibration, the Gold06 dataset is
plagued with statistical inhomogeneities which are possibly due to
systematic errors. Given the fact that the more recent data (SNLS
and HST) are statistically consistent with each other and
homogeneous, it is highly probable that the possible source of
systematic errors lies within the earlier data and in particular
in the HZSST subset.

{\bf Numerical Analysis:} The mathematica files and the datafile
used in the numerical analysis of this work may be found at
http://leandros.physics.uoi.gr/gold06/gold06.htm or may be sent by
e-mail upon request.

\section*{Acknowledgements}
This work was supported by the European Research and Training
Network MRTPN-CT-2006 035863-1 (UniverseNet). SN acknowledges
support from the Greek State Scholarships Foundation (I.K.Y.).

\clearpage
\newpage

\nopagebreak[0]
\section{Appendix}
\begin{table}[h!]
\begin{center}
\caption{The Gold06 dataset with its subsets. The \label{table4}}
\begin{tabular}{ccccc}
\hline
\hline\\

\textbf{SN} & \textbf{$z$}  &  \textbf{$\mu_0$} & \textbf{$\sigma_{\mu_0}$} &     \textbf{Subsample} \\
\hline \hline \\

SN03D1au    \hspace{7pt} &    0.504    \hspace{7pt} &    42.61    \hspace{7pt} &    0.17    \hspace{7pt} &    SNLS    \hspace{7pt} \\
SN03D1aw    \hspace{7pt} &    0.582    \hspace{7pt} &    43.07    \hspace{7pt} &    0.17    \hspace{7pt} &    SNLS    \hspace{7pt} \\
SN03D1ax    \hspace{7pt} &    0.496    \hspace{7pt} &    42.36    \hspace{7pt} &    0.17    \hspace{7pt} &    SNLS    \hspace{7pt} \\
SN03D1cm    \hspace{7pt} &    0.870    \hspace{7pt} &    44.28    \hspace{7pt} &    0.34    \hspace{7pt} &    SNLS    \hspace{7pt} \\
SN03D1co    \hspace{7pt} &    0.679    \hspace{7pt} &    43.58    \hspace{7pt} &    0.19    \hspace{7pt} &    SNLS    \hspace{7pt} \\
SN03D1fc    \hspace{7pt} &    0.331    \hspace{7pt} &    41.13    \hspace{7pt} &    0.17    \hspace{7pt} &    SNLS    \hspace{7pt} \\
SN03D1fl    \hspace{7pt} &    0.688    \hspace{7pt} &    43.23    \hspace{7pt} &    0.17    \hspace{7pt} &    SNLS    \hspace{7pt} \\
SN03D1fq    \hspace{7pt} &    0.800    \hspace{7pt} &    43.67    \hspace{7pt} &    0.19    \hspace{7pt} &    SNLS    \hspace{7pt} \\
SN03D3af    \hspace{7pt} &    0.532    \hspace{7pt} &    42.78    \hspace{7pt} &    0.18    \hspace{7pt} &    SNLS    \hspace{7pt} \\
SN03D3aw    \hspace{7pt} &    0.449    \hspace{7pt} &    42.05    \hspace{7pt} &    0.17    \hspace{7pt} &    SNLS    \hspace{7pt} \\
SN03D3ay    \hspace{7pt} &    0.371    \hspace{7pt} &    41.67    \hspace{7pt} &    0.17    \hspace{7pt} &    SNLS    \hspace{7pt} \\
SN03D3bh    \hspace{7pt} &    0.249    \hspace{7pt} &    40.76    \hspace{7pt} &    0.17    \hspace{7pt} &    SNLS    \hspace{7pt} \\
SN03D3cc    \hspace{7pt} &    0.463    \hspace{7pt} &    42.27    \hspace{7pt} &    0.17    \hspace{7pt} &    SNLS    \hspace{7pt} \\
SN03D3cd    \hspace{7pt} &    0.461    \hspace{7pt} &    42.22    \hspace{7pt} &    0.17    \hspace{7pt} &    SNLS    \hspace{7pt} \\
SN03D4ag    \hspace{7pt} &    0.285    \hspace{7pt} &    40.92    \hspace{7pt} &    0.17    \hspace{7pt} &    SNLS    \hspace{7pt} \\
SN03D4at    \hspace{7pt} &    0.633    \hspace{7pt} &    43.32    \hspace{7pt} &    0.18    \hspace{7pt} &    SNLS    \hspace{7pt} \\
SN03D4cx    \hspace{7pt} &    0.949    \hspace{7pt} &    43.69    \hspace{7pt} &    0.32    \hspace{7pt} &    SNLS    \hspace{7pt} \\
SN03D4cz    \hspace{7pt} &    0.695    \hspace{7pt} &    43.21    \hspace{7pt} &    0.19    \hspace{7pt} &    SNLS    \hspace{7pt} \\
SN03D4dh    \hspace{7pt} &    0.627    \hspace{7pt} &    42.93    \hspace{7pt} &    0.17    \hspace{7pt} &    SNLS    \hspace{7pt} \\
SN03D4di    \hspace{7pt} &    0.905    \hspace{7pt} &    43.89    \hspace{7pt} &    0.30    \hspace{7pt} &    SNLS    \hspace{7pt} \\
SN03D4dy    \hspace{7pt} &    0.604    \hspace{7pt} &    42.70    \hspace{7pt} &    0.17    \hspace{7pt} &    SNLS    \hspace{7pt} \\
SN03D4fd    \hspace{7pt} &    0.791    \hspace{7pt} &    43.54    \hspace{7pt} &    0.18    \hspace{7pt} &    SNLS    \hspace{7pt} \\
SN03D4gg    \hspace{7pt} &    0.592    \hspace{7pt} &    42.75    \hspace{7pt} &    0.19    \hspace{7pt} &    SNLS    \hspace{7pt} \\
SN03D4gl    \hspace{7pt} &    0.571    \hspace{7pt} &    42.65    \hspace{7pt} &    0.18    \hspace{7pt} &    SNLS    \hspace{7pt} \\
SN04D1ag    \hspace{7pt} &    0.557    \hspace{7pt} &    42.70    \hspace{7pt} &    0.17    \hspace{7pt} &    SNLS    \hspace{7pt} \\
SN04D2cf    \hspace{7pt} &    0.369    \hspace{7pt} &    41.67    \hspace{7pt} &    0.17    \hspace{7pt} &    SNLS    \hspace{7pt} \\
SN04D2fp    \hspace{7pt} &    0.415    \hspace{7pt} &    41.96    \hspace{7pt} &    0.17    \hspace{7pt} &    SNLS    \hspace{7pt} \\
SN04D2fs    \hspace{7pt} &    0.357    \hspace{7pt} &    41.63    \hspace{7pt} &    0.17    \hspace{7pt} &    SNLS    \hspace{7pt} \\
SN04D2gb    \hspace{7pt} &    0.430    \hspace{7pt} &    41.96    \hspace{7pt} &    0.17    \hspace{7pt} &    SNLS    \hspace{7pt} \\
SN04D2gp    \hspace{7pt} &    0.707    \hspace{7pt} &    43.42    \hspace{7pt} &    0.21    \hspace{7pt} &    SNLS    \hspace{7pt} \\
SN04D3co    \hspace{7pt} &    0.620    \hspace{7pt} &    43.21    \hspace{7pt} &    0.18    \hspace{7pt} &    SNLS    \hspace{7pt} \\
SN04D3cy    \hspace{7pt} &    0.643    \hspace{7pt} &    43.21    \hspace{7pt} &    0.18    \hspace{7pt} &    SNLS    \hspace{7pt} \\
SN04D3df    \hspace{7pt} &    0.470    \hspace{7pt} &    42.45    \hspace{7pt} &    0.17    \hspace{7pt} &    SNLS    \hspace{7pt} \\
SN04D3do    \hspace{7pt} &    0.610    \hspace{7pt} &    42.98    \hspace{7pt} &    0.17    \hspace{7pt} &    SNLS    \hspace{7pt} \\
SN04D3ez    \hspace{7pt} &    0.263    \hspace{7pt} &    40.87    \hspace{7pt} &    0.17    \hspace{7pt} &    SNLS    \hspace{7pt} \\
SN04D3fk    \hspace{7pt} &    0.358    \hspace{7pt} &    41.66    \hspace{7pt} &    0.17    \hspace{7pt} &    SNLS    \hspace{7pt} \\
SN04D3fq    \hspace{7pt} &    0.730    \hspace{7pt} &    43.47    \hspace{7pt} &    0.18    \hspace{7pt} &    SNLS    \hspace{7pt} \\
SN04D3hn    \hspace{7pt} &    0.552    \hspace{7pt} &    42.65    \hspace{7pt} &    0.17    \hspace{7pt} &    SNLS    \hspace{7pt} \\
SN04D3kr    \hspace{7pt} &    0.337    \hspace{7pt} &    41.44    \hspace{7pt} &    0.17    \hspace{7pt} &    SNLS    \hspace{7pt} \\
SN04D3lu    \hspace{7pt} &    0.822    \hspace{7pt} &    43.73    \hspace{7pt} &    0.27    \hspace{7pt} &    SNLS    \hspace{7pt} \\
SN04D3ml    \hspace{7pt} &    0.950    \hspace{7pt} &    44.14    \hspace{7pt} &    0.31    \hspace{7pt} &    SNLS    \hspace{7pt} \\
SN04D3nh    \hspace{7pt} &    0.340    \hspace{7pt} &    41.51    \hspace{7pt} &    0.17    \hspace{7pt} &    SNLS    \hspace{7pt} \\
SN04D3oe    \hspace{7pt} &    0.756    \hspace{7pt} &    43.64    \hspace{7pt} &    0.17    \hspace{7pt} &    SNLS    \hspace{7pt} \\
SN04D4an    \hspace{7pt} &    0.613    \hspace{7pt} &    43.15    \hspace{7pt} &    0.18    \hspace{7pt} &    SNLS    \hspace{7pt} \\
SN04D4bq    \hspace{7pt} &    0.550    \hspace{7pt} &    42.67    \hspace{7pt} &    0.17    \hspace{7pt} &    SNLS    \hspace{7pt} \\
SN04D4dm    \hspace{7pt} &    0.811    \hspace{7pt} &    44.13    \hspace{7pt} &    0.31    \hspace{7pt} &    SNLS    \hspace{7pt} \\
SN04D4dw    \hspace{7pt} &    0.961    \hspace{7pt} &    44.18    \hspace{7pt} &    0.33    \hspace{7pt} &    SNLS    \hspace{7pt} \\
1997ff      \hspace{7pt} &    1.755    \hspace{7pt} &    45.35    \hspace{7pt} &    0.35    \hspace{7pt} &    HST    \hspace{7pt} \\
2002dc      \hspace{7pt} &    0.475    \hspace{7pt} &    42.24    \hspace{7pt} &    0.20    \hspace{7pt} &    HST    \hspace{7pt} \\
2002dd      \hspace{7pt} &    0.950    \hspace{7pt} &    43.98    \hspace{7pt} &    0.34    \hspace{7pt} &    HST    \hspace{7pt} \\
2003eq      \hspace{7pt} &    0.840    \hspace{7pt} &    43.67    \hspace{7pt} &    0.21    \hspace{7pt} &    HST    \hspace{7pt} \\
2003es      \hspace{7pt} &    0.954    \hspace{7pt} &    44.30    \hspace{7pt} &    0.27    \hspace{7pt} &    HST    \hspace{7pt} \\
2003eb      \hspace{7pt} &    0.900    \hspace{7pt} &    43.64    \hspace{7pt} &    0.25    \hspace{7pt} &    HST    \hspace{7pt} \\
\hline \hline
\end{tabular}
\end{center}
\end{table}

\begin{table} [h!]
\vspace{47pt}
\begin{center}
six outliers of the HZSST subset are denoted by a $^*$. \\
\begin{tabular}{ccccc}
\hline
\hline\\

\textbf{SN} & \textbf{$z$}  &  \textbf{$\mu_0$} & \textbf{$\sigma_{\mu_0}$} &     \textbf{Subsample} \\
\hline \hline \\
2003XX      \hspace{7pt} &    0.935    \hspace{7pt} &    43.97    \hspace{7pt} &    0.29    \hspace{7pt} &    HST    \hspace{7pt} \\
2003bd      \hspace{7pt} &    0.670    \hspace{7pt} &    43.19    \hspace{7pt} &    0.24    \hspace{7pt} &    HST    \hspace{7pt} \\
2002kd      \hspace{7pt} &    0.735    \hspace{7pt} &    43.14    \hspace{7pt} &    0.19    \hspace{7pt} &    HST    \hspace{7pt} \\
2003be      \hspace{7pt} &    0.640    \hspace{7pt} &    43.01    \hspace{7pt} &    0.25    \hspace{7pt} &    HST    \hspace{7pt} \\
2003dy      \hspace{7pt} &    1.340    \hspace{7pt} &    44.92    \hspace{7pt} &    0.31    \hspace{7pt} &    HST    \hspace{7pt} \\
2002ki      \hspace{7pt} &    1.140    \hspace{7pt} &    44.71    \hspace{7pt} &    0.29    \hspace{7pt} &    HST    \hspace{7pt} \\
2002hp      \hspace{7pt} &    1.305    \hspace{7pt} &    44.51    \hspace{7pt} &    0.30    \hspace{7pt} &    HST    \hspace{7pt} \\
2002fw      \hspace{7pt} &    1.300    \hspace{7pt} &    45.06    \hspace{7pt} &    0.20    \hspace{7pt} &    HST    \hspace{7pt} \\
HST04Pat    \hspace{7pt} &    0.970    \hspace{7pt} &    44.67    \hspace{7pt} &    0.36    \hspace{7pt} &    HST    \hspace{7pt} \\
HST04Mcg    \hspace{7pt} &    1.370    \hspace{7pt} &    45.23    \hspace{7pt} &    0.25    \hspace{7pt} &    HST    \hspace{7pt} \\
HST05Fer    \hspace{7pt} &    1.020    \hspace{7pt} &    43.99    \hspace{7pt} &    0.27    \hspace{7pt} &    HST    \hspace{7pt} \\
HST05Koe    \hspace{7pt} &    1.230    \hspace{7pt} &    45.17    \hspace{7pt} &    0.23    \hspace{7pt} &    HST    \hspace{7pt} \\
HST04Gre    \hspace{7pt} &    1.140    \hspace{7pt} &    44.44    \hspace{7pt} &    0.31    \hspace{7pt} &    HST    \hspace{7pt} \\
HST04Omb    \hspace{7pt} &    0.975    \hspace{7pt} &    44.21    \hspace{7pt} &    0.26    \hspace{7pt} &    HST    \hspace{7pt} \\
HST05Lan    \hspace{7pt} &    1.230    \hspace{7pt} &    44.97    \hspace{7pt} &    0.20    \hspace{7pt} &    HST    \hspace{7pt} \\
HST04Tha    \hspace{7pt} &    0.954    \hspace{7pt} &    43.85    \hspace{7pt} &    0.27    \hspace{7pt} &    HST    \hspace{7pt} \\
HST04Rak    \hspace{7pt} &    0.740    \hspace{7pt} &    43.38    \hspace{7pt} &    0.22    \hspace{7pt} &    HST    \hspace{7pt} \\
HST04Yow    \hspace{7pt} &    0.460    \hspace{7pt} &    42.23    \hspace{7pt} &    0.32    \hspace{7pt} &    HST    \hspace{7pt} \\
HST04Man    \hspace{7pt} &    0.854    \hspace{7pt} &    43.96    \hspace{7pt} &    0.29    \hspace{7pt} &    HST    \hspace{7pt} \\
HST05Spo    \hspace{7pt} &    0.839    \hspace{7pt} &    43.45    \hspace{7pt} &    0.20    \hspace{7pt} &    HST    \hspace{7pt} \\
HST04Eag    \hspace{7pt} &    1.020    \hspace{7pt} &    44.52    \hspace{7pt} &    0.19    \hspace{7pt} &    HST    \hspace{7pt} \\
HST05Gab    \hspace{7pt} &    1.120    \hspace{7pt} &    44.67    \hspace{7pt} &    0.18    \hspace{7pt} &    HST    \hspace{7pt} \\
HST05Str    \hspace{7pt} &    1.010    \hspace{7pt} &    44.77    \hspace{7pt} &    0.19    \hspace{7pt} &    HST    \hspace{7pt} \\
HST04Sas    \hspace{7pt} &    1.390    \hspace{7pt} &    44.90    \hspace{7pt} &    0.19    \hspace{7pt} &    HST    \hspace{7pt} \\
SN95K    \hspace{7pt} &    0.478    \hspace{7pt} &    42.48    \hspace{7pt} &    0.23    \hspace{7pt} &    HZSST    \hspace{7pt} \\
SN96E    \hspace{7pt} &    0.425    \hspace{7pt} &    41.69    \hspace{7pt} &    0.40    \hspace{7pt} &    HZSST    \hspace{7pt} \\
SN96H    \hspace{7pt} &    0.620    \hspace{7pt} &    43.11    \hspace{7pt} &    0.28    \hspace{7pt} &    HZSST    \hspace{7pt} \\
SN96I    \hspace{7pt} &    0.570    \hspace{7pt} &    42.80    \hspace{7pt} &    0.25    \hspace{7pt} &    HZSST    \hspace{7pt} \\
SN96J    \hspace{7pt} &    0.300    \hspace{7pt} &    41.01    \hspace{7pt} &    0.25    \hspace{7pt} &    HZSST    \hspace{7pt} \\
SN96K    \hspace{7pt} &    0.380    \hspace{7pt} &    42.02    \hspace{7pt} &    0.22    \hspace{7pt} &    HZSST    \hspace{7pt} \\
SN96U    \hspace{7pt} &    0.430    \hspace{7pt} &    42.33    \hspace{7pt} &    0.34    \hspace{7pt} &    HZSST    \hspace{7pt} \\
SN97as    \hspace{7pt} &    0.508    \hspace{7pt} &    42.19    \hspace{7pt} &    0.35    \hspace{7pt} &    HZSST    \hspace{7pt} \\
SN97bb    \hspace{7pt} &    0.518    \hspace{7pt} &    42.83    \hspace{7pt} &    0.31    \hspace{7pt} &    HZSST    \hspace{7pt} \\
SN97bj    \hspace{7pt} &    0.334    \hspace{7pt} &    40.92    \hspace{7pt} &    0.30    \hspace{7pt} &    HZSST    \hspace{7pt} \\
SN97ce    \hspace{7pt} &    0.440    \hspace{7pt} &    42.07    \hspace{7pt} &    0.19    \hspace{7pt} &    HZSST    \hspace{7pt} \\
SN97cj    \hspace{7pt} &    0.500    \hspace{7pt} &    42.73    \hspace{7pt} &    0.20    \hspace{7pt} &    HZSST    \hspace{7pt} \\
SN98ac    \hspace{7pt} &    0.460    \hspace{7pt} &    41.81    \hspace{7pt} &    0.40    \hspace{7pt} &    HZSST    \hspace{7pt} \\
SN98M    \hspace{7pt} &    0.630    \hspace{7pt} &    43.26    \hspace{7pt} &    0.37    \hspace{7pt} &    HZSST    \hspace{7pt} \\
SN98J    \hspace{7pt} &    0.828    \hspace{7pt} &    43.59    \hspace{7pt} &    0.61    \hspace{7pt} &    HZSST    \hspace{7pt} \\
SN99Q2$^*$    \hspace{7pt} &    0.459    \hspace{7pt} &    42.67    \hspace{7pt} &    0.22    \hspace{7pt} &    HZSST    \hspace{7pt} \\
SN99U2    \hspace{7pt} &    0.511    \hspace{7pt} &    42.83    \hspace{7pt} &    0.21    \hspace{7pt} &    HZSST    \hspace{7pt} \\
SN99S$^*$    \hspace{7pt} &    0.474    \hspace{7pt} &    42.81    \hspace{7pt} &    0.22    \hspace{7pt} &    HZSST    \hspace{7pt} \\
SN99N    \hspace{7pt} &    0.537    \hspace{7pt} &    42.85    \hspace{7pt} &    0.41    \hspace{7pt} &    HZSST    \hspace{7pt} \\
SN99fn    \hspace{7pt} &    0.477    \hspace{7pt} &    42.38    \hspace{7pt} &    0.21    \hspace{7pt} &    HZSST    \hspace{7pt} \\
SN99ff    \hspace{7pt} &    0.455    \hspace{7pt} &    42.29    \hspace{7pt} &    0.28    \hspace{7pt} &    HZSST    \hspace{7pt} \\
SN99fj    \hspace{7pt} &    0.815    \hspace{7pt} &    43.75    \hspace{7pt} &    0.33    \hspace{7pt} &    HZSST    \hspace{7pt} \\
SN99fm    \hspace{7pt} &    0.949    \hspace{7pt} &    44.00    \hspace{7pt} &    0.24    \hspace{7pt} &    HZSST    \hspace{7pt} \\
SN99fk    \hspace{7pt} &    1.056    \hspace{7pt} &    44.35    \hspace{7pt} &    0.23    \hspace{7pt} &    HZSST    \hspace{7pt} \\
SN99fw    \hspace{7pt} &    0.278    \hspace{7pt} &    41.01    \hspace{7pt} &    0.41    \hspace{7pt} &    HZSST    \hspace{7pt} \\
SN99fv$^*$    \hspace{7pt} &    1.199    \hspace{7pt} &    44.19    \hspace{7pt} &    0.34    \hspace{7pt} &    HZSST    \hspace{7pt} \\
SN00ec$^*$    \hspace{7pt} &    0.470    \hspace{7pt} &    42.76    \hspace{7pt} &    0.21    \hspace{7pt} &    HZSST    \hspace{7pt} \\
SN00dz    \hspace{7pt} &    0.500    \hspace{7pt} &    42.74    \hspace{7pt} &    0.24    \hspace{7pt} &    HZSST    \hspace{7pt} \\
SN00eg    \hspace{7pt} &    0.540    \hspace{7pt} &    41.96    \hspace{7pt} &    0.41    \hspace{7pt} &    HZSST    \hspace{7pt} \\
\hline \hline
\end{tabular}
\end{center}
\end{table}

\clearpage
\newpage

\begin{table} [h!]
\begin{center}
TABLE IV continued \\
\begin{tabular}{ccccc}
\hline
\hline\\

\textbf{SN} & \textbf{$z$}  &  \textbf{$\mu_0$} & \textbf{$\sigma_{\mu_0}$} &     \textbf{Subsample} \\
\hline \hline \\

SN00ee$^*$    \hspace{7pt} &    0.470    \hspace{7pt} &    42.73    \hspace{7pt} &    0.23    \hspace{7pt} &    HZSST    \hspace{7pt} \\
SN00eh    \hspace{7pt} &    0.490    \hspace{7pt} &    42.40    \hspace{7pt} &    0.25    \hspace{7pt} &    HZSST    \hspace{7pt} \\
SN01jh    \hspace{7pt} &    0.884    \hspace{7pt} &    44.22    \hspace{7pt} &    0.19    \hspace{7pt} &    HZSST    \hspace{7pt} \\
SN01hu    \hspace{7pt} &    0.882    \hspace{7pt} &    43.89    \hspace{7pt} &    0.30    \hspace{7pt} &    HZSST    \hspace{7pt} \\
SN01iy    \hspace{7pt} &    0.570    \hspace{7pt} &    42.87    \hspace{7pt} &    0.31    \hspace{7pt} &    HZSST    \hspace{7pt} \\
SN01jp    \hspace{7pt} &    0.528    \hspace{7pt} &    42.76    \hspace{7pt} &    0.25    \hspace{7pt} &    HZSST    \hspace{7pt} \\
SN01fo$^*$    \hspace{7pt} &    0.771    \hspace{7pt} &    43.12    \hspace{7pt} &    0.17    \hspace{7pt} &    HZSST    \hspace{7pt} \\
SN01hs    \hspace{7pt} &    0.832    \hspace{7pt} &    43.55    \hspace{7pt} &    0.29    \hspace{7pt} &    HZSST    \hspace{7pt} \\
SN01hx    \hspace{7pt} &    0.798    \hspace{7pt} &    43.88    \hspace{7pt} &    0.31    \hspace{7pt} &    HZSST    \hspace{7pt} \\
SN01hy    \hspace{7pt} &    0.811    \hspace{7pt} &    43.97    \hspace{7pt} &    0.35    \hspace{7pt} &    HZSST    \hspace{7pt} \\
SN01jf    \hspace{7pt} &    0.815    \hspace{7pt} &    44.09    \hspace{7pt} &    0.28    \hspace{7pt} &    HZSST    \hspace{7pt} \\
SN01jm    \hspace{7pt} &    0.977    \hspace{7pt} &    43.91    \hspace{7pt} &    0.26    \hspace{7pt} &    HZSST    \hspace{7pt} \\
SN95aw    \hspace{7pt} &    0.400    \hspace{7pt} &    42.04    \hspace{7pt} &    0.19    \hspace{7pt} &    SCP    \hspace{7pt} \\
SN95ax    \hspace{7pt} &    0.615    \hspace{7pt} &    42.85    \hspace{7pt} &    0.23    \hspace{7pt} &    SCP    \hspace{7pt} \\
SN95ay    \hspace{7pt} &    0.480    \hspace{7pt} &    42.37    \hspace{7pt} &    0.20    \hspace{7pt} &    SCP    \hspace{7pt} \\
SN95az    \hspace{7pt} &    0.450    \hspace{7pt} &    42.13    \hspace{7pt} &    0.21    \hspace{7pt} &    SCP    \hspace{7pt} \\
SN95ba    \hspace{7pt} &    0.388    \hspace{7pt} &    42.07    \hspace{7pt} &    0.19    \hspace{7pt} &    SCP    \hspace{7pt} \\
SN96ci    \hspace{7pt} &    0.495    \hspace{7pt} &    42.25    \hspace{7pt} &    0.19    \hspace{7pt} &    SCP    \hspace{7pt} \\
SN96cl    \hspace{7pt} &    0.828    \hspace{7pt} &    43.96    \hspace{7pt} &    0.46    \hspace{7pt} &    SCP    \hspace{7pt} \\
SN97eq    \hspace{7pt} &    0.538    \hspace{7pt} &    42.66    \hspace{7pt} &    0.18    \hspace{7pt} &    SCP    \hspace{7pt} \\
SN97ek    \hspace{7pt} &    0.860    \hspace{7pt} &    44.03    \hspace{7pt} &    0.30    \hspace{7pt} &    SCP    \hspace{7pt} \\
SN97ez    \hspace{7pt} &    0.778    \hspace{7pt} &    43.81    \hspace{7pt} &    0.35    \hspace{7pt} &    SCP    \hspace{7pt} \\
SN97F    \hspace{7pt} &    0.580    \hspace{7pt} &    43.04    \hspace{7pt} &    0.21    \hspace{7pt} &    SCP    \hspace{7pt} \\
SN97H    \hspace{7pt} &    0.526    \hspace{7pt} &    42.56    \hspace{7pt} &    0.18    \hspace{7pt} &    SCP    \hspace{7pt} \\
SN97I    \hspace{7pt} &    0.172    \hspace{7pt} &    39.79    \hspace{7pt} &    0.18    \hspace{7pt} &    SCP    \hspace{7pt} \\
SN97N    \hspace{7pt} &    0.180    \hspace{7pt} &    39.98    \hspace{7pt} &    0.18    \hspace{7pt} &    SCP    \hspace{7pt} \\
SN97P    \hspace{7pt} &    0.472    \hspace{7pt} &    42.46    \hspace{7pt} &    0.19    \hspace{7pt} &    SCP    \hspace{7pt} \\
SN97Q    \hspace{7pt} &    0.430    \hspace{7pt} &    41.99    \hspace{7pt} &    0.18    \hspace{7pt} &    SCP    \hspace{7pt} \\
SN97R    \hspace{7pt} &    0.657    \hspace{7pt} &    43.27    \hspace{7pt} &    0.20    \hspace{7pt} &    SCP    \hspace{7pt} \\
SN97ac    \hspace{7pt} &    0.320    \hspace{7pt} &    41.45    \hspace{7pt} &    0.18    \hspace{7pt} &    SCP    \hspace{7pt} \\
SN97af    \hspace{7pt} &    0.579    \hspace{7pt} &    42.86    \hspace{7pt} &    0.19    \hspace{7pt} &    SCP    \hspace{7pt} \\
SN97ai    \hspace{7pt} &    0.450    \hspace{7pt} &    42.10    \hspace{7pt} &    0.23    \hspace{7pt} &    SCP    \hspace{7pt} \\
SN97aj    \hspace{7pt} &    0.581    \hspace{7pt} &    42.63    \hspace{7pt} &    0.19    \hspace{7pt} &    SCP    \hspace{7pt} \\
SN97am    \hspace{7pt} &    0.416    \hspace{7pt} &    42.10    \hspace{7pt} &    0.19    \hspace{7pt} &    SCP    \hspace{7pt} \\
SN97ap    \hspace{7pt} &    0.830    \hspace{7pt} &    43.85    \hspace{7pt} &    0.19    \hspace{7pt} &    SCP    \hspace{7pt} \\
SN98ba    \hspace{7pt} &    0.430    \hspace{7pt} &    42.36    \hspace{7pt} &    0.25    \hspace{7pt} &    SCP    \hspace{7pt} \\
SN98bi    \hspace{7pt} &    0.740    \hspace{7pt} &    43.35    \hspace{7pt} &    0.30    \hspace{7pt} &    SCP    \hspace{7pt} \\
SN00fr    \hspace{7pt} &    0.543    \hspace{7pt} &    42.67    \hspace{7pt} &    0.19    \hspace{7pt} &    SCP    \hspace{7pt} \\
\hline \hline
\end{tabular}
\end{center}
\end{table}

\begin{table} [h!]
\vspace{5pt}
\begin{center}
TABLE IV continued \\
\begin{tabular}{ccccc}
\hline
\hline\\

\textbf{SN} & \textbf{$z$}  &  \textbf{$\mu_0$} & \textbf{$\sigma_{\mu_0}$} &     \textbf{Subsample} \\
\hline \hline \\
SN92bs    \hspace{7pt} &    0.063    \hspace{7pt} &    37.67    \hspace{7pt} &    0.19    \hspace{7pt} &    LR    \hspace{7pt} \\
SN94M    \hspace{7pt} &    0.024    \hspace{7pt} &    35.09    \hspace{7pt} &    0.22    \hspace{7pt} &    LR    \hspace{7pt} \\
SN94T    \hspace{7pt} &    0.036    \hspace{7pt} &    36.01    \hspace{7pt} &    0.21    \hspace{7pt} &    LR    \hspace{7pt} \\
SN97dg    \hspace{7pt} &    0.029    \hspace{7pt} &    36.13    \hspace{7pt} &    0.21    \hspace{7pt} &    LR    \hspace{7pt} \\
SN00bk    \hspace{7pt} &    0.026    \hspace{7pt} &    35.35    \hspace{7pt} &    0.23    \hspace{7pt} &    LR    \hspace{7pt} \\
SN98cs    \hspace{7pt} &    0.032    \hspace{7pt} &    36.08    \hspace{7pt} &    0.20    \hspace{7pt} &    LR    \hspace{7pt} \\
SN00cf    \hspace{7pt} &    0.036    \hspace{7pt} &    36.39    \hspace{7pt} &    0.19    \hspace{7pt} &    LR    \hspace{7pt} \\
SN98dx    \hspace{7pt} &    0.053    \hspace{7pt} &    36.95    \hspace{7pt} &    0.19    \hspace{7pt} &    LR    \hspace{7pt} \\
SN99gp    \hspace{7pt} &    0.026    \hspace{7pt} &    35.57    \hspace{7pt} &    0.21    \hspace{7pt} &    LR    \hspace{7pt} \\
SN99X    \hspace{7pt} &    0.025    \hspace{7pt} &    35.40    \hspace{7pt} &    0.22    \hspace{7pt} &    LR    \hspace{7pt} \\
SN99cc    \hspace{7pt} &    0.031    \hspace{7pt} &    35.84    \hspace{7pt} &    0.21    \hspace{7pt} &    LR    \hspace{7pt} \\
SN94Q    \hspace{7pt} &    0.029    \hspace{7pt} &    35.70    \hspace{7pt} &    0.21    \hspace{7pt} &    LR    \hspace{7pt} \\
SN95ac    \hspace{7pt} &    0.049    \hspace{7pt} &    36.55    \hspace{7pt} &    0.20    \hspace{7pt} &    LR    \hspace{7pt} \\
SN96bl    \hspace{7pt} &    0.034    \hspace{7pt} &    36.19    \hspace{7pt} &    0.20    \hspace{7pt} &    LR    \hspace{7pt} \\
SN90O    \hspace{7pt} &    0.030    \hspace{7pt} &    35.90    \hspace{7pt} &    0.21    \hspace{7pt} &    LR    \hspace{7pt} \\
SN96C    \hspace{7pt} &    0.027    \hspace{7pt} &    35.90    \hspace{7pt} &    0.21    \hspace{7pt} &    LR    \hspace{7pt} \\
SN96ab    \hspace{7pt} &    0.124    \hspace{7pt} &    39.19    \hspace{7pt} &    0.22    \hspace{7pt} &    LR    \hspace{7pt} \\
SN99ef    \hspace{7pt} &    0.038    \hspace{7pt} &    36.67    \hspace{7pt} &    0.19    \hspace{7pt} &    LR    \hspace{7pt} \\
SN92J    \hspace{7pt} &    0.046    \hspace{7pt} &    36.35    \hspace{7pt} &    0.21    \hspace{7pt} &    LR    \hspace{7pt} \\
SN92bk    \hspace{7pt} &    0.058    \hspace{7pt} &    37.13    \hspace{7pt} &    0.19    \hspace{7pt} &    LR    \hspace{7pt} \\
SN92bp    \hspace{7pt} &    0.079    \hspace{7pt} &    37.94    \hspace{7pt} &    0.18    \hspace{7pt} &    LR    \hspace{7pt} \\
SN92br    \hspace{7pt} &    0.088    \hspace{7pt} &    38.07    \hspace{7pt} &    0.28    \hspace{7pt} &    LR    \hspace{7pt} \\
SN93H    \hspace{7pt} &    0.025    \hspace{7pt} &    35.09    \hspace{7pt} &    0.22    \hspace{7pt} &    LR    \hspace{7pt} \\
SN93ah    \hspace{7pt} &    0.028    \hspace{7pt} &    35.53    \hspace{7pt} &    0.22    \hspace{7pt} &    LR    \hspace{7pt} \\
SN90T    \hspace{7pt} &    0.040    \hspace{7pt} &    36.38    \hspace{7pt} &    0.20    \hspace{7pt} &    LR    \hspace{7pt} \\
SN90af    \hspace{7pt} &    0.050    \hspace{7pt} &    36.84    \hspace{7pt} &    0.22    \hspace{7pt} &    LR    \hspace{7pt} \\
SN91U    \hspace{7pt} &    0.033    \hspace{7pt} &    35.53    \hspace{7pt} &    0.21    \hspace{7pt} &    LR    \hspace{7pt} \\
SN91S    \hspace{7pt} &    0.056    \hspace{7pt} &    37.31    \hspace{7pt} &    0.19    \hspace{7pt} &    LR    \hspace{7pt} \\
SN92P    \hspace{7pt} &    0.026    \hspace{7pt} &    35.63    \hspace{7pt} &    0.22    \hspace{7pt} &    LR    \hspace{7pt} \\
SN92bg    \hspace{7pt} &    0.036    \hspace{7pt} &    36.17    \hspace{7pt} &    0.20    \hspace{7pt} &    LR    \hspace{7pt} \\
SN92bl    \hspace{7pt} &    0.043    \hspace{7pt} &    36.52    \hspace{7pt} &    0.19    \hspace{7pt} &    LR    \hspace{7pt} \\
SN92bh    \hspace{7pt} &    0.045    \hspace{7pt} &    36.99    \hspace{7pt} &    0.18    \hspace{7pt} &    LR    \hspace{7pt} \\
SN92au    \hspace{7pt} &    0.061    \hspace{7pt} &    37.31    \hspace{7pt} &    0.22    \hspace{7pt} &    LR    \hspace{7pt} \\
SN92ae    \hspace{7pt} &    0.075    \hspace{7pt} &    37.77    \hspace{7pt} &    0.19    \hspace{7pt} &    LR    \hspace{7pt} \\
SN92aq    \hspace{7pt} &    0.101    \hspace{7pt} &    38.70    \hspace{7pt} &    0.20    \hspace{7pt} &    LR    \hspace{7pt} \\
SN93ag    \hspace{7pt} &    0.050    \hspace{7pt} &    37.07    \hspace{7pt} &    0.19    \hspace{7pt} &    LR    \hspace{7pt} \\
SN93O    \hspace{7pt} &    0.052    \hspace{7pt} &    37.16    \hspace{7pt} &    0.18    \hspace{7pt} &    LR    \hspace{7pt} \\
SN93B    \hspace{7pt} &    0.071    \hspace{7pt} &    37.78    \hspace{7pt} &    0.19    \hspace{7pt} &    LR    \hspace{7pt} \\
\hline \hline
\end{tabular}
\end{center}
\end{table}
\clearpage
\end{document}